\documentclass[openacc]{rstransa}%%%%where rstrans is the template name
\jname{rsta} 
\Journal{{\em Phil.\ Trans.\ R. Soc.\ A\ }  }

\usepackage{braket}
\usepackage{bbm}
\newcommand{\ketbra}[1]{|#1\rangle\langle #1|}
\newcommand{\deq}[1]{ \begin{align}#1\end{align}}
\newcommand{\bS}{\mathbf{S}}
\newcommand{\Tr}{\mathrm{Tr}}
\newcommand{\cH}{\mathcal{H}}
\newcommand{\cI}{\mathcal{I}}
\newcommand{\cL}{\mathcal{L}}
\newcommand{\R}{\mathbbm{R}}
\newcommand{\da}{\dagger}
\newcommand{\id}{{\rm id}}
\newcommand{\nn}{\nonumber}
\newcommand{\ph}{\phi}
\newcommand{\rh}{\rho}
\newcommand{\Eq}[1]{Eq.~(\ref{eq:#1})}
 \newcommand{\bx}{{\bf x}}
\newcommand{\by}{{\bf y}}
\newcommand{\eq}[1]{(\ref{eq:#1})}
\newcommand{\bA}{{\bf A}}
\newcommand{\bE}{{\bf E}}
\newcommand{\bM}{{\bf M}}
\newcommand{\cK}{{\cal K}}
\titlehead{Research}

\title{Incorporating episodic memory into quantum  models of judgment and decision}

\author{%%%% Author details
Jerome R. Busemeyer$^{1}$, Masanao Ozawa$^{2,3,4,5}$, Emmanuel M. Pothos$^{6}$, and Naotsugu Tsuchiya$^{7,8,9}$
}

\address{\noindent
{\footnotesize
$^{1}$Psychological Brain Sciences, Indiana University, Bloomington In 47405 USA\\
$^{2}$G.S. Informatics, Nagoya University, Nagoya 464-8601 Japan\\
$^{3}$Kinugasa R.O., Ritsumeikan University, Kyoto 603-8577 Japan\\
$^{4}$CMSAI, Chubu University, Kasugai, Aichi 487-8501 Japan\\
$^{5}$RIKEN TRIP FQSP, iTHEMS, Wako, Saitama 351-0198 Japan\\
$^{6}$City, University of London, Social Sciences Building, 32-38 Whiskin Street, London EC1R 0JD\\
$^{7}$School of Psychological Sciences, Faculty of Medicine, Nursing and Health Science, 
Monash University, Victoria, Australia\\
$^{8}$Center  for  Information  and  Neural  Networks  (CiNet),  
National  Institute  of  Information  and Communications Technology (NICT), Osaka, Japan\\
$^{9}$Advanced  Telecommunications  Research  Computational  Neuroscience  Laboratories,  Kyoto, Japan\\
}
}

\subject{quantum cognition}

\keywords{quantum cognition, system plus environment models, instrument measurements}

\corres{Jerome Busemeyer\\
\email{jbusemey@iu.edu}}

\begin{document}
%%%% Abstract text to be placed here %%%%%%%%%%%%
\begin{abstract}
\sloppy
An important challenge for quantum theories of cognition and decision concerns the incorporation of memory for recently made judgments and their effects on later judgments. First, we review a general approach to measurement based on system plus environment representations of states and measurement instruments. 
These more general measurement models provide ways to incorporate effects of recent judgments on later judgments.  Then we compare three different measurement models that are based on these more general measurement operations to a puzzling collection of question order effect findings.  
\end{abstract}
%%%%%%%%%%%%%%%%%%%%%%%%%%%

%%%%%%%%%% Insert the texts which can accomdate on firstpage in the tag "fmtext" %%%%%
\begin{fmtext}
\section{Introduction}
Probability theory has been fruitfully employed across many areas of cognitive science as a way to formalize models that allow precise and falsifiable predictions. The relevance of probability theory is unsurprising, since, whatever else, part of what the mind is trying to accomplish must be trying to process and extrapolate from statistical structure in the environment. 
\end{fmtext}
\maketitle
%%%%%%%%%%%%%%% End of first page %%%%%%%%%%%%%%%%%%%%%

The most common approach is based on Bayesian probability theory (see, e.g. \cite{1}).  More recently, some researchers have been exploring the use of quantum probability theory in cognitive science  (see, e.g., \cite{2,3}).  Quantum models use the probability rules from quantum theory to model human behavior, but without the physics. These models were introduced to explain what seemed to be, from a classical perspective, reasoning fallacies and irrational decisions \cite{4}. 

An important challenge for both theories concerns the incorporation of episodic memories. How can recently experienced events be included in the models? 

\section{Definition of episodic memory }

Episodic memory refers to the memory of recently experienced events, which differs from semantic memory of their meaning \cite{5}. For example, remembering what was said in the morning news is episodic, but understanding its meaning is semantic.  Here, we focus solely on episodic memory; we briefly discuss semantic memory at the end of the paper.

Research on episodic memory has a long history in psychology, beginning as early as 1885 with Hermann Ebbinghaus studying 
recalling lists of non-sense syllables. Subsequently, a very large literature of facts has accumulated on this topic, with much of the research following Ebbinghous by studying recalling word lists, lists of numbers, or lists of pictures. This research has produced a great body of empirical facts, including, for example, serial position effects on episodic memory \cite{6}. Alongside these many empirical facts, many prominent cognitive models have been developed to account for these facts (see \cite{5}  for a review). Some initial work on quantum models of episodic memory has  begun with the development of models designed to account for a puzzling finding called the memory over-distribution effect (see, e.g., \cite{7,8}). However, this paper addresses a different issue concerning the effects of episodic memory on later judgments and decisions.

\section{Incorporating episodic memory into judgments}

Both Bayesian and quantum models are well suited for updating a belief state based on recently experienced events from environmental sources providing new information. For example, upon hearing news that the government passed some tax bill, an investor, whom we will call Anita, can update her beliefs about what might be good investments in the future. However, theories differ with regard to the effects of recent self-made judgments.  For example, before hearing the news, a client might ask Anita what she predicts will happen.  By making a prediction, Anita gains no new information, and so according to a Bayesian model,  her belief state will remain intact after her prediction; in any case, there is no requirement in Bayesian theory for a belief state to change as a result of a self-measurement. However, according to a quantum cognition model, this judgment could act as a measurement of Anita's beliefs that produces a reduction or ``collapse'' of the quantum state to a state that is consistent with the observed outcome.  Hereafter, we are mainly concerned with ways to build models of episodic memory of events, either informative or judgmental, for quantum models.

\subsection{Born rule}
Quantum theory represents a system $\bS$ under investigation by a Hilbert space $\cH$ with
inner product $(\xi,\eta)$ for all $\xi,\eta\in\cH$.\footnote{We follow the physics convention that the inner product 
$(\xi,\eta)$ is linear in $\eta$ and conjugate linear in $\xi$.}
We assume $\cH$ to be finite dimensional.
The state of the system $\bS$ is represented by a unit vector, called a state vector, in $\cH$.
An observable quantity (or, observable, for brevity) of the system $\bS$ is represented by a self-adjoint operator on $\cH$.\footnote{
Note that an operator $A$ on $\cH$ is called self-adjoint if $A^{\dagger}=A$,
where the adjoint $A^{\dagger}$ of $A$ is defined by $(\xi,A^{\dagger}\eta)=(A\xi,\eta)$ for all 
$\xi,\eta\in\cH$. So, $A^{\dagger}=A$ implies $(\psi,A\psi)\in\R$ for all $\psi\in\cH$, since $(\psi,A\psi)=(A\psi,\psi)=(\psi,A\psi)^{*}$.
It follows that all the eigenvalues $a$ of $A$ are real numbers, since $a=(\psi,A\psi)$ for a normalized eigenvector $\psi$ such that $A\psi=a\psi$.} 
For any observable $A$ and a real number $a\in\R$, we define its spectral projection $P^{A}(a)$ as the projection onto the subspace
$\{\psi\in\cH\mid A\psi=a\psi\}$ of $\cH$.
Then, $a\in\R$ is an eigenvalue of $A$ if and only if $P^{A}(a)\not=0$.

Quantum theory has two rules for the results of a measurement of an observable,
the \textit{Born rule} and the \textit{state update rule}. 
The Born rule states that if an observable $A$ is measured in a state $\psi$, 
we observe the outcome $A=a$ with probability $p(A=a\|\psi)=(\psi,P^{A}(a)\psi)$.
From the above, we see that any possible outcome $a\in\R$ is one of the eigenvalues of $A$, since 
$p(A=a\|\psi)>0$ implies $P^{A}(a)\not=0$.

\subsection{Von Neumann--L\"{u}ders projection postulate}
As a consequence of the Schr\"{o}dinger equation, the state of a system $\bS$ that does not interact with any external systems
changes in time as the {\em rule of unitary evolution}: There is a unitary operator $U(t,t')$ for any time interval $[t,t']$ 
such that the state $\psi(t)$ of the system $\bS$ at time $t$ changes to
the state $\psi(t')$ at time $t'$ as $\psi(t')=U(t',t)\psi(t)$.

However, we need another rule, the state update rule, to describe the state change during a measurement, 
since any measurement needs a non-trivial interaction with an environment.
Conventionally, the state update has been modeled in quantum theory, as well as quantum cognition, by projecting the state 
of the system onto a subspace that represents the observed outcome. 
Thus, the conventional state update rule requires that if an observable $A$ of the system $\bS$ is measured in a state $\psi$
and the outcome $A=a$ is observed, then the state of the system $\bS$ be updated, or ``collapse'',  to the post-measurement state
$\psi_{\{A=a\}}=\frac{P^{A}(a)\psi}{\sqrt{p(A=a\|\psi)}}$.\footnote{Note that $\sqrt{p(A=a\|\psi)}$ is the normalization factor, 
since $\sqrt{p(A=a\|\psi)}=\|P^{A}(a)\psi\|$.}
This state update rule is called the {\em von Neumann-L\"{u}ders projection postulate} or the {\em projection postulate} in short.

\subsection{Measurement models for quantum cognition}
In quantum cognition, question-answer process is modeled as a measurement of an observable $A$ with 2 different eigenvalues 0 and 1,
where the answers ``yes'' and ``no'' are encoded to 1 and 0, respectively;  such an observable is always a projection, which satisfies
$A=A^{\dagger}=A^2$, and we have $P^{A}(1)=A$ and $P^{A}(0)=I-A$, where $I$ is the identity operator on $\cH$.   

For example, consider a simple 2-dimensional space used to represent Anita's state  
$\psi_{A}$ with respect to a pair of basis vectors : 
$\psi_A=c_{1}\left|relaxed\right\rangle +c_{2}\left|not\;relaxed\right\rangle $. If her friend asks her how she is feeling, and she decides that she is feeling relaxed, then Anita's state is projected on the basis vector $\left|relaxed\right\rangle $ and normalized. 
Here, we encode the state $\ket{relaxed}$ and $\ket{not\;relaxed}$ to the eigenstates 
of the observable $A$ with $P^{A}(1)=\ketbra{relaxed}$ and  $P^{A}(0)=\ketbra{not\;relaxed}$.

This kind of projection onto a single basis vector was originally proposed by von Neumann \cite{9} to represent the impact of measurement on a quantum state, and it has often been employed in simple 2-dimensional quantum cognition models (see, e.g., \cite{10}).
More precisely, von Neumann \cite{9} introduced the {\em repeatability hypothesis}, which states: “If a physical quantity is measured twice in succession in a system, then we get the same value each time” \cite[p.~335]{9}. This hypothesis was formulated as a general principle extracted from the Compton--Simons experiment.

Equivalent formulations were also given by Dirac \cite[p.~36]{11} and Sch\"{o}dinger \cite[\S 8]{12}. 
This requires to update the state of the system to be an {\em arbitrary} eigenstate corresponding 
to the outcome of the measurement. So, the repeatability hypothesis is 
more general than the projection postulate.
However, if the eigensubspace is one-dimensional, then the eigenstate is unique (up to a phase factor), 
so that in this case the measurement projects the state onto the subspace corresponding to the outcome.

Anita's state may not be limited to only 2 dimensions and she may be able to experience many more than just 2 states of relaxation. For example, she could feel only a little bit relaxed or she could feel very relaxed.  Suppose she can experience a finer grain scale of relaxation with 101 ordered states ranging along a lattice from very much not relaxed to undecided to very much relaxed. She may report relaxed whenever her state is greater than undecided. In that case, her state would initially be in a superposition lying in a $N=101$ dimensional space, and her answer ``relax'' would project her state onto a 50 dimensional subspace spanned by all the states greater than undecided, again with normalization.  
 This kind of projection onto a multi-dimensional subspace was later proposed by L\"{u}ders \cite{13} (so it is sometimes called the L\"{u}ders  rule) and it has also been frequently employed in quantum cognition (see, e.g., \cite{14}). 

Now suppose Anita is asked another question about her mood and she answers that she is in a good mood. Define $P_{relax}$ as the projection for the ``relaxed'' answer, and define $P_{mood}$ as the projection for the ``good mood'' answer.  Suppose these projections do not commute so that $P_{relax}  P_{mood} \ne P_{mood}  P_{relax}$.  Noncommutativity of projections provides a basis for explaining question order effects. Then, according to the uncertainty principle, collapsing to the answer for the question about mood will interfere with memory about the previous answer to the question about relaxation.  Additional incompatible questions could   continue to interfere with earlier memory about the relaxation answer.  
In general, if (and only if) two questions corresponds to commuting projections, 
those question does not destroy the episodic memory about the previous report in the sense that all the reports are logically consistent
 \cite{15}. 
In short, even the von Neumann--L\"{u}ders projection rule fails to provide an adequate account of episodic memory compatible 
with question order effects. 
To overcome this limitation, the next two sections provide  modern quantum measurement models that can retain full information and memory for the outcomes of previous measurements.

\subsection{Density operators}
The description of the state of a system by a state vector leads 
to two different types of combinations of states, superposition and mixture. 
A state $\psi$ is called a superposition of states $\psi_1, \ldots, \psi_n$ if there are complex 
numbers $\alpha_1,\ldots, \alpha_n$ such that $\psi=\sum_{j=1}^{n}\alpha_j\psi_j$.  
The superposition of states remains described by a state vector.

On the other hand, mixture requires a new mathematical representation. 
We say that the system $\bS$ is in the mixture of states  $\psi_1, \ldots, \psi_n$ 
with probabilities $p_1,\ldots,p_n$ (or, in the mixture $(\psi_j,p_i)$, for brevity),
if  the system $\bS$ is in the state $\psi_j$ with probability $p_j$.
To represent the mixture as a linear combination of states, 
let $\ketbra{\psi}$ be the the projection onto the one-dimensional subspace spanned by $\psi$.\footnote{In general, the operator $\ket{\xi}\bra{\eta}$
is defined as $\ket{\xi}\bra{\eta}\psi=(\eta,\psi)\xi$, where $\xi,\eta,\psi\in\cH$.}
Then, we have $p(A=a\|\psi)=\Tr[P^{A}(a) \ketbra{\psi}]$.
Now, we introduce the operator $\rho=\sum_{j}p_{j}\ketbra{\psi_j}$
called the density operator of the mixture $(\psi_j,p_j)$.
Then, the density operator $\rho$ naturally extends the Born rule to mixtures
as $p(A=a\|(\psi_j,p_j))=\sum_j p_j p(A=a\|\psi_j)=\Tr[P^{A}(a)\rho]$ for any observable $A$.

The operator $\rho$ is a positive operator with unit trace, i.e., $(\psi,\rho\psi)\ge 0$ for all
 $\psi\in\cH$ and $\Tr[\rho]=1$.  We generally call any positive operator with unit trace as a density operator.
  Any density operator arises from at least one mixture $(\psi_j,p_j)$, since the spectral decomposition $\rho=\sum_{j}p_{j}\ketbra{\psi_j}$ provides a mixture $(\psi_j,p_j)$ corresponding to $\rho$.
 Extending the the notion of state, we postulate that the state of the system $\bS$ 
 is represented by a density operator $\rho$
 with the new Born rule $p(A=a\|\rho)=\Tr[P^{A}(a)\rho]$.
To distinguish the original notion, 
 if $\rho=\ketbra{\psi}$ we say that the system is in the pure state $\psi$. 
 We say that the system is in the mixture of states $\rho_1,\ldots,\rho_n$ with probabilities
$p_1,\ldots,p_n$ if $\rho=\sum_{j}p_j\rho_j$.
The mixture of states remains described by a density operator.
 
  The rule of unitary evolution is extended as the relation 
 $\rho(t')= U(t',t)\rho(t) U(t',t)^{\dagger}$ if the system $\bS$ is in the state 
 $\rho(t)$ at time $t$ and $\rho(t')$ at time $t'$.\footnote{The relation follows
 from $\ketbra{\psi(t')}=U(t',t)\ketbra{\psi(t)}U(t',t)^{\dagger}$ and 
 $\rho(t')=\sum_{j}p_j \ketbra{\psi(t')}$ if 
 $\rho(t)=\sum_{j}p_j \ketbra{\psi(t)}$.
 }
 The von Neumaann-L\"{u}ders projection postulate is extended to 
 mixtures:
 If an observable $A$ is measured in a state $\rho$
and the outcome $A=a$ is observed, then the state of the system $\bS$ is updated to the post-measurement state
$\rho_{\{A=a\}}=\frac{P^{A}(a)\rho P^{A}(a)}{p(A=a\|\rho)}$; this extends the original formulation, since
$\rho_{\{A=a\}}=\ketbra{\psi_{\{A=a\}}}$ if $\rho=\ketbra{\psi}$.
\footnote{Note that $p(A=a\|\rho)$ is the normalization factor, 
since $p(A=a\|\psi)=\Tr[P^{A}(a)\rho P^{A}(a)]$.}

One of the main reasons for using a density operator is that it allows for an additional kind of uncertainty. 
A superposition $\psi=\sum_{j}\alpha_j\psi_j$ allows an individual to be indefinite about outcomes $E=1$ or $E=0$, where $E=\ketbra{\psi_j}$,
with respect to some basis $\{\psi_j\}$. However, a pure state assumes that the theorist has maximal knowledge of the system state \cite{12}. Usually the theorist does not have maximal knowledge and instead there may be some uncertainty in the state. In this case, the system could be in $L$ possible states
$\psi_j$ with $j=1,\ldots, L$, and each possible state has a probability $p_j$. If we are modeling a group of people, then $p_j$ represents the probability of including a person in state 
$\psi_j$.
If we are  modeling an individual, then $p_j$ may represent the probability that the individual is prepared in state $\psi_j$.
The density operator provides a general way to allow for this additional uncertainty about  the exact state.  

A projection has the properties $P_{A}(a)=P_{A}(a)^{\dagger}=P_{A}(a)^{2}$ so that the probability of repeatedly 
observing the outcome $A=a$ immediately after observing the outcome $A=a$ is 1, i.e., 
$p(A=a\|\rho_{\{A=a\}})=\frac{\Tr[P^{A}(a)P^{A}(a)\rho P^{A}(a)]}{\Tr[P^{A}(a)\rho]}=1$.
Thus, the projection postulate implies the repeatability hypothesis.

\subsection{Quantum instruments}
As we have seen in the preceding subsections, the results of an measurement of an observable $A$ are represented by 
the Born rule, $p(A=a\|\rho)=\Tr[P^{A}(a)\rho]$, and the von Neumann-L\"{u}ders projection postulate 
$\rho\to \rho_{\{A=a\}}=\frac{P^{A}(a)\rho P^{A}(a)}{\Tr[P^{A}(a)\rho]}$.
However, Davies and Lewis \cite{16} proposed to abandon the von Neumann-L\"{u}ders projection postulate (or more generally, the repeatability hypothesis) to provide a more flexible measurement theory.\footnote{``One of the crucial notions is that of repeatability which we show is implicitly assumed in most of the axiomatic treatments of quantum mechanics, but whose
abandonment leads to a much more flexible approach to measurement theory.''
\cite[p.~239]{16}.}
For this purpose, Davies and Lewis \cite{16} introduced the mathematical notion of ``instrument'' as follows.
Let $\cI=\{\cI(x)\}_{x\in\R}$ be a family of linear transformations $\cI(x)$ on the space $\cL(\cH)$ of operators on $\cH$.
We shall write $\cI(S)=\sum_{x\in S}\cI(x)$ for any subset $S\subseteq\R$, hereafter.
Then,  $\cI$  is called
an instrument if (i) $\cI(x)$ is positive for all $x\in\R$, i.e., $\cI(x)\rho\ge 0$ if $\rho\ge 0$, and (ii) $\cI(\R)$ is trace-preserving, i.e,  $\Tr[\cI(\R)\rho]=\Tr[\rho]$ for all $\rho\in\cL(\cH)$.\footnote{A linear transformation $T$ on the space $\cL(\cH)$
of operators is called a superoperator.
A superoperator is said to be positive if it transforms a positive operator to a
positive operator, where an operator $\rho\in\cL(\cH)$ is positive if $(\psi,A\psi)\ge 0$ for any $\psi\in\cH$. A positive superoperator is called
an operation if it does not increase the trace of density operators, i.e., $\Tr[T\rho]\le 1$ for every density operator $\rho$, so that an instrument $\cI$ is a family of operations $\cI(x)$ such that $\cI(\R)$ is trace-preserving.}

Consider a measuring apparatus $\bA(\bx)$ with the classical output variable  $\bx$.
The statistical properties of the apparatus $\bA(\bx)$ 
are specified  by (i) the probability $p(\bx=x\|\rh)$ of the outcome $\bx=x$ in any
pre-measurement state $\rho$, and
(ii) the state update $\rho\to \rh_{\{\bx=x\}}$, where  $\rh_{\{\bx=x\}}$ is the post-measurement state to the pre-measurement state $\rho$
for the outcome $\bx=x$.
Then, Davies and Lewis introduced the following postulate ({\em Davies-Lewis postulate}): 
For every apparatus $\bA(\bx)$ there exist  a unique instrument
$\cI$ satisfying  (i)
$p(\bx=x\|\rh)=\Tr[\cI(x)\rh]$, and (ii) 
$\rho\to\rho_{\{\bx=x\}}=\frac{\cI(x)\rh}{\Tr[\cI(x)\rh]}$.
In what follows, we shall identify the apparatus $\bA(\bx)$ with its associate instrument $\cI$.
Note that conditions (i) and (ii) above are equivalent to the single relation 
\deq{\label{eq:DL3}\cI(x)\rho=p(\bx=x\|\rh)\rh_{\{\bx=x\}}.}

A family $\Pi=\{\Pi(x)\}_{x\in\R}$ of positive operators $\Pi(x)$ is called a POVM (probability operator-valued
measure) if $\sum_{x\in\R}\Pi(x)=I$, where $I\in\cL(\cH)$ is the identity operator on $\cH$.
For any superoperator $T$ on $\cL(\cH)$, its dual $T^{*}$ is a superoperator on  $\cL(\cH)$ satisfying
$\Tr[A(T\rho)]=\Tr[(T^{*}A)\rho]$ for all $A,\rho\in\cL(\cH)$.
For any instrument $\cI$, the family of operators $\Pi(x)$ defined by $\Pi(x)=\cI(x)^{*}I$ is a POVM and called the POVM associated with $\cI$, which 
satisfies $p(\bx=x\|\rh)=\Tr[\Pi(x)\rho]$ for any $\rho\in\cL(\cH)$.\footnote{
Here, $I\in\cL(\cH)$ denoted the identity operator on $\cH$.
Note that $\Pi(x)\rho$ denotes the product of
two operators $\Pi(x)$ and $\rho$, whereas $\cI(x)\rh$ denotes the operator obtained by applying the superoperator $\cI(x)$ to the operator $\rho$.
So, the dual $\cI(x)^{*}$ of $\cI(x)$ satisfies $p(\bx=x\|\rh)=\Tr[\cI(x)\rho]=\Tr[I(\cI(x)\rho)]
=\Tr[(\cI(x)^{*}I)\rh]=\Tr[\Pi(x)\rh]$.} 
The probability of the outcome $\bx=x$ of the instrument 
$\cI$ in a state $\rho$ satisfies $p(\bx=x\|\rh)=\Tr[\Pi(x)\rho]$.   

We say that an instrument $\cI$ measures an observable $A$ 
if $p(\bx=x\|\rh)=\Tr[P^{A}(x)\rho]$. Then, we have 
$\Tr[\cI(x)\rho]=\Tr[P^{A}(x)\rho]$  for any pre-measurement state $\rho$ 
and the POVM $\Pi$ associated with $\cI$ satisfies $\Pi=P^{A}$.
An instrument $\cI$ is called a projective instrument for an observable $A$ if $\cI(x)\rho=P^{A}(x)\rho P^{A}(x)$.
The von Neumann-L\"{u}ders projection postulate requires that
the statistical properties of any apparatus 
be described by a projective instrument, whereas the Davies-Lewis postulate 
abandon this restriction to allow any instrument to describe the statistical 
properties of possible measurements.

Let $\bA(\bx)$ and $\bA(\by)$ be two measuring apparatuses with 
instruments $\cI_\bx$ and $\cI_\by$, respectively.
Consider the sequential measurements by $\bA(\bx)$ and $\bA(\by)$ in this order
in the pre-measurement state $\rho$.\footnote{In sequential measurements,
we assume that the time just before the next measurement equals the time 
just after the preceding measurement, or there is no external intervention
between two consecutive measurements. }
Then, the sequential probability distribution $p(\bx=x,\by=y\|\rh)$ of
$\bx$ and $\by$ is given by
\deq{\label{eq:101121}
p(\bx=x,\by=y\|\rh)=
p(\by=y\|\rh_{\{\bx=x\}})
p(\bx=x\|\rh)
=\Tr[\cI_\by(y)\cI_\bx(x)\rho].
}
By induction, if we make sequential measurements using the apparatuses $\bA(\bx_1),\ldots,\bA(\bx_n)$ in this order
with instruments $\cI_{1},\ldots,\cI_{n}$, we have
\deq{ p(\bx_1=x_1,\ldots,\bx_n=x_n\|\rh)
=\Tr[\cI_n(x_n)\cdots\cI_1(x_1)\rho].}

From \Eq{101121} the above sequential probability distribution 
has the following property ({\em Mixing law for sequential outcome probability}):
\label{P:ML}
For any sequential measurements carried out by apparatuses
$\bA(\bx)$ and $\bA(\by)$ in this order, the sequential probability distribution
$p(\bx=x,\by=y\|\rh)$ of outcomes $\bx$ and $\by$
is an affine function in $\rh$, i.e., 
\deq{p(\bx=x,\by=y\|\rh)=
p_1p(\bx=x,\by=y\|\rh_1)+
p_2p(\bx=x,\by=y\|\rh_2),}
if $\rho=p_1\rho_1+p_2\rho_2$.
It was shown that this natural property of the sequential outcome probability 
for sequential measurements is indeed logically equivalent to the Davies-Lewis postulate \cite[Theorem 2]{17} (see also \cite[Theorem 2.3]{18}, \cite[\S 2.3]{19}).
So, the Davies-Lewis postulate is considered as a necessary condition to describe
all the physically realizable measuring apparatuses.

Quantum instruments were introduced in quantum theory quite a while ago to generalize the concept of quantum measurement (see \cite{16,20,21}).  Recently, they have been used in quantum cognition (see, \cite{22}, for a recent review on this measurement theory). Like a collapse, quantum instruments change the state in a non-unitary manner that loses information when an event is observed. Unlike a collapse, they do not necessarily reduce the state by a projection onto a subspace, as done with the L\"{u}ders rule. Instruments allow ``weak'' or noisy measurements so that there is only some probability of repeating a response immediately after the application of the instrument. Also instruments can produce modifications of the state of the system in a way that provides some imperfect and possibly noisy episodic memory for the observed event. 

\subsection{Measuring processes}
Von Neumann \cite[Chapter VI]{9} showed that a projective instrument can be described by the interaction between the system and an environment,
and the subsequent observation of the meter 
in the environment to determine the outcome of the measurement.
Generalizing von Neumann's model to arbitrary physically realizable measurements,
a general model of measuring process is introduced in \cite{20}:
A {\em measuring process} (or, {\em indirect measurement model})
for the system $\bS$ described by a Hilbert space $\cH$
is defined as a quadruple $\bM=(\cK,\xi,U,M)$ consisting of a Hilbert space
$\cK$, a state vector $\xi$ in $\cK$, a unitary operator $U$ on the tensor
produce Hilbert space $\cH\otimes\cK$, and a self-adjoint operator $M$
on $\cK$.

The measuring process $\bM$ models the following process of a measurement
with an apparatus $\bA(\bx)$.
The measurement is carried out by the interaction between the measured system $\bS$ 
described by a Hilbert space $\cH$ 
and an environment $\bE$ described by a Hilbert space $\cK$.
The environment is in a fixed pure state $\xi$ just before the measurement, 
whereas the measured system $\bS$ is in an arbitrary state $\rho$.
We assume that the system $\bS$ and the environment $\bE$ interact only in the time interval during the measurement \cite[\S VIII]{17};\footnote{This requirement is necessary for the partial trace over the environment determines the system state. } in particular, they are statistically independent just before the measurement.
So, the composite system $\bS+\bE$ is initially in the state $\rho\otimes\ketbra{\xi}$.
According to the rule of unitary evolution, the time evolution of the composite system $\bS+\bE$ during the measuring interaction is 
represented by a unitary operator $U$.
So, the composite system $\bS+\bE$ is in the state $U(\rho\otimes\ketbra{\xi})U^{\da}$
just after the measurement.
Then, the  outcome $\bx=x$ is obtained 
by measuring the observable $M$, called the meter observable, in the environment $\bE$.
To determine the instrument $\cI$ of this measuring process, suppose that an observable $B$ of the system $\bS$ 
is observed just after the measurement. Then, from \Eq{DL3} we obtain
\deq{
\Tr\left[P^{B}(b)\left[\cI(x)\rho\right]\right]
&=p(\bx=x,B=b\|\rho)\nn\\
&=\Tr\left[\left(P^{B}(b)\otimes P^{M}(x)\right)U(\psi\otimes\ketbra{\xi})U^{\da}\right]\nn\\
&=\Tr\left[P^{B}(b)\Tr_{\cK}\left[\left(I\otimes P^{M}(x)\right)U(\psi\otimes\ketbra{\xi})U^{\da}\right]\right],
}
where $\Tr_{\cK}$ stands for the partial trace on the Hilbert space
$\cK$.
Since $B$ was arbitrary, we determine the instrument $\cI$ of the measuring process 
$\bM=(\cK,\xi,U,M)$ as
\deq{\label{eq:instrument_MP}
\cI(x)\rh=\Tr_{\cK}\left[\left(1\otimes P^{M}(x)\right)
U(\rh\otimes\ketbra{\xi})U^{\da}\right].
}
In fact, \Eq{instrument_MP} determines a family of positive superoperators $\{\cI(x)\}_{x\in \R}$
such that $\cI(\R)$ is trace-preserving.\footnote{The last condition follows from 
$\Tr[\cI(\R)\rho]=\Tr\left[\left(1\otimes P^{M}(\R)\right)
U(\rh\otimes\ketbra{\xi})U^{\da}\right]=\Tr[U(\rh\otimes\ketbra{\xi})U^{\da}]=
\Tr[\rh\otimes\ketbra{\xi}]=\Tr[\rho]\Tr[\ketbra{\xi}]=\Tr[\rho]$.}
See \cite{20} for the detailed derivation.\footnote{
Note that, in contrast to a wide-spread popular view,
we never appeal to the projection postulate for the measurement of the 
meter observable $M$ in deriving the state update rule \eq{instrument_MP}.
In fact, \Eq{instrument_MP} holds even in the case where
the measurement of the meter observable $M$ is not a projective
measurement; for a detailed discussion on this point see 
\cite{20,23,21,24,17,25}.}
Thus, the statistical properties of the measuring process $\bM$ are given by 
\deq{
p(\bx=x\|\rh)
&=\Tr \left[\left(1\otimes P^{M}(x)\right)
U(\rh\otimes\ketbra{\xi})U^{\da}\right],\\
\rh_{\{\bx=x\}}
&=\dfrac{\Tr_{\cK}\left[\left(1\otimes P^{M}(x)\right)
U(\rh\otimes\ketbra{\xi})U^{\da}\right]}
{\Tr \left[\left(1\otimes P^{M}(x)\right)
U(\rh\otimes\ketbra{\xi})U^{\da}\right]}.
}

In quantum cognition, the system refers to a representation of the belief or values of the individual under investigation (e.g., Anita's opinions).  
The measurement could either be an event  that is observed to occur in the environment (e.g., Anita hearing the news) or answers to questions that the person is being asked (e.g., Anita's prediction). 
In the case where if the system $\bS$ is in a pure state $\psi$, the system plus environment state is 
a vector contained in an $N\times M$ dimensional Hilbert space formed by 
the tensor product of 
an $N$ dimensional system space and an $M$ dimensional environment space.  
Initially, before the measurement occurs, 
or before any questions are asked, the state is formed by a tensor product, $\Psi=\psi\otimes\xi$
of an initial system state $\psi$
and an initial environment state $\xi$.
Initially, the system state is uncorrelated with the environment state. 
The measurement 
is now represented by a unitary operator, $U$, that entangles and correlates the system and environment parts of the state to produce a final measured state, $U(\psi\otimes\xi)$.
In quantum cognition, the probability of answering a question with the response  $\bx=1$ (yes)  or $\bx=0$ (no)
is obtained from observing the meter observable represented by the projection $P_E=I\otimes P^{M}(1)$
in the state just after the measurement, 
as we have the relation $M=P^{M}(1)$ if $M$ has only two eigenvalues $1$ and $0$.
Note that the projection $M=P^{M}(1)$  represents a property of the environment 
that indicates the outcome of the measurement.
Instead of the single projection $M=P^{M}(1)$ , 
a family of mutually orthogonal projections $P^{M}(x)$, 
where $x$ varies over a scale $S=\{-n,\ldots,0,\ldots,n\}$, often 
represents a family of possible outcomes from the measurement, and
called the  \textit{pointer position} of the measurement.

\subsection{Complete positivity}
Given an instrument $\cI$ for the system with a Hilbert space $\cH$, 
a measuring process $\bM=(\cK,\xi,U,M)$ satisfying \Eq{instrument_MP}
is called a realization of the instrument $\cI$.
An instrument $\cI$ is called realizable if there exists its realization.
Is every instrument realizable, or what instruments are realizable?
To solve this problem we introduce the following definition.
A superoperator $T$ on $\cL(\cH)$ is called {\it completely positive} 
if it satisfies
\deq{
\sum_{i,j=1}^{n}B_{i}T(A_{i}A^{\da}_{j})B^{\da}_{j}\ge 0
}
for any finite sequences  $A_{1},\ldots,A_{n}, B_{1},\ldots,B_{n}\in\cL(\cH)$.%
\footnote{
The above definition is equivalent to requiring that
the superoperator $T\otimes\id$ on $\cL(\cH)\otimes \cL(\cK)$
defined by $T\otimes\id(X\otimes Y)=T(X)\otimes Y$ for all $X\in\cL(\cH)$ and 
$Y\in\cL(\cK)$ be a positive superoperator on  $\cL(\cH)\otimes \cL(\cK)$ for every 
finite dimensional Hilbert space $\cK$.}
An instrument for $\cH$ is called a {\it completely positive (CP)
instrument} if $\cI(x)$ is completely positive for every $x\in\R$.
It can be seen from \Eq{instrument_MP} that every realizable instrument is completely 
positive, and moreover it is known that an instrument is realizable if and only if it is completely positive \cite{20}.
Thus, completely positive instruments mathematically characterize all the
possible quantum measurements consistent with quantum theory (see \cite{18} for more
detailed discussions).

The transpose map on $\cL(\cH)$ relative to an orthonormal basis $\{\ph_j\}$
of $\cH$ is a superoperator defined by $T(\ket{\ph_j}\bra{\ph_k}) = \ket{\ph_k}\bra{\ph_j}$ for all $j,k$ is a trace-preserving positive superoperator but not completely positive  \cite{26}.
For any observable $A$, we can define an instrument $\cI$ by 
$\cI_A(x)\rho=T(P^{A}{(x)}\rho P^{A}{(x)})$ for any $\rho\in\cL(\cH)$.
Then, $\cI_A$ measures $A$ but is not completely positive.
So, for any observable $A$, there is an $A$-measuring instrument that is not realizable \cite{19,20,21}. 

On the other hand, it is known that every completely positive superoperator $T$ on 
$\cL(\cH)$ there exists a family of operators $\{M_j\}$ such that $T\rho=\sum_j M_j \rho M_j^{\dagger}$
\cite{27}.
So, for any completely positive instrument $\cI$ there exists a family of operators
$\{M_{xj}\}_{xj}$  in $\cL(\cH)$ with $x\in\R$ and $j=1,2,\ldots$, called the measurement operators for $\cI$, such that $\cI(x)\rho=\sum_{j}M_{xj}\rho M_{xj}^{\dagger}$ \cite{22}.
In this case, the POVM $\Pi$ associated with $\cI$ satisfies 
$\Pi(x)=\sum_{j}M_{xj}^{\dagger}M_{xj}$.

\section{Application to question order phenomena}

An important problem concerning measurement theory in quantum cognition relates to findings from three different experimental paradigms \cite{28}. The first is the AA (and BB) paradigm, in which the same question A is asked twice, back to back. It is commonly assumed that all participants will certainly repeat the first answer on the second occasion, however, note that there is very scant evidence that this assumption is actually empirically valid.
The second is the AB vs. BA, question order  paradigm, in which questions A and B are asked in different orders. Question order effects have frequently been empirically observed and when they do, they generally satisfy a prediction from quantum cognition called the $QQ$ equality \cite{29}: 
\[
QQ = [p(A=y,B=n) + p(A=n,B=y)] - [p(B=y,A=n) + p(B=n,A=y)] = 0.
\]
 The third finding is from the so-called ABA (and BAB) paradigm, in which question A is followed by question B, which is finally followed by asking question A again. Some participants change their answers to question A when it is repeated, but others simply repeat the original answer to question A \cite{30}. The problem is how to account for all three findings.  The 
 projective measurement model, based on non-commuting observables and the von Neumann-L\"{u}ders projection rule, can account for the findings from first two paradigms
 \cite{31,29}. The problem arises with the ABA paradigm for participants who produce question order effects (QOE) on the AB vs. BA paradigm \cite{28}. 
The effect combining AA, BB, ABA, and BAB paradigms is called the response replicability effect (RRE).
 A complete analysis of the compatibility of QOE and RRE in the projective model was given in \cite{15}
 and it was shown that if a projective model shows RRE with a state $\psi$, then the two projections representing
 the questions A and B  must commute in that state, so that projective models do not show the combination 
 of  QOE and RRE with any state. Note that a projective model is specified by two projections corresponding to the 
 pair of questions A and B and a state vector $\psi$ with which all statistics of the sequential outcomes are derived.
Thus, to realize both QOE and RRE, we need more general quantum models other than projective models.  

The first work to successfully address all three finding was by Ozawa and Khrennikov  (denoted the OK account) \cite{32} using instrument theory. A later account for all three findings was proposed by Busemeyer  (denoted the BB account) \cite{33,34}  using the system plus environment (indirect measurement) approach. Below we briefly review the basic ideas of each theory with a new account proposed first in the present paper (denoted the OB account), which reformulates the BB account using instrument theory. Later we compare them.

\subsection{Instrument (OK) Model}

The OK account starts out by assuming that the Hilbert space is composed of a tensor product of a system plus environment space. The system space is a tensor product of a 2 dimensional space used to represent a person's belief or opinion about question A, and another 2 dimensional space used to represent the person's belief on question B, and a 3 dimensional space representing the type of an individual (type just refers to the way of belief update in answering a question). The system space is spanned by the set of basis vectors  $\Omega=\left\{ \left|0\right\rangle ,\left|1\right\rangle \right\} \times\left\{ \left|0\right\rangle ,\left|1\right\rangle \right\} \times\left\{ \left|0\right\rangle,\left|1\right\rangle,\left|2\right\rangle\right\} $, whereby the first and second parts represent the binary values (e.g., false, true) for the belief of the person on each question, with the third part representing the type of the person's belief update. For example, (0,1,2) indicates a person of type 2 who believes the answer to A is no and the answer to B is yes. This forms a $12$ dimensional Hilbert space.

The environment is represented by another 2 - dimensional space used to represent a person's response for question A and another 2 dimensional space used to represent the person's response to question B; both spaces are spanned by the basis vectors  $\left\{ \left|0\right\rangle,\left|1\right\rangle\right\} \times\left\{ \left|0\right\rangle,\left|1\right\rangle\right\} $.  This forms a $4$ dimensional space. According to the OK model, this environmental space represents the physically recorded responses (i.e., the answers typed into a computer), rather than mentally recorded episodic memories of the response. 

The tensor product, system plus environment, state is therefore a $48$ dimensional space. For example, one of the basis vectors is $\ket{102}\ket{00}$
indicating a person of type 2 (third coordinate) with an initial belief of yes to question A (first coordinate) and no to question B (second coordinate); the final two coordinates are the initial state of the environment. 

Each person's initial mental state can be identified as exactly one of the 12 basis states in $\Omega$ defining the system space (see section 5 in \cite{32}).\footnote{The OK model allows for superposition states, but the model with superposition states eventually reduces to a model that is identical to one assuming each person starts out in exactly one of the basis states. See appendix D in \cite{32}.} A mixed state is formed by assigning a probability, denoted $\mu_{\alpha,\beta,\gamma}$,  that the person starts in each one of the 12 basis states indexed by $\alpha,\beta,\gamma$.  In other words, the model can be described as one in which every individual starts out with a clear belief, but different individuals start with different beliefs. The initial environment state is fixed to be equal to $\left|00\right\rangle $.  

A unitary operator, either $U_A$ or $U_B$, is applied depending on whether the focus is on question A, or B respectively.  The behaviors of these unitary operators depend on the type of person. Type 0 people simply copy the values on the belief states directly onto the response states, without any back action updating the system state for both unitary operators. 

Type 1 people also copy the value on the belief states directly onto the response states, but there is also a back action that updates the system state as follows. For question A, if the values on the belief state agree, then the value on belief B is updated  to disagree with A and the belief of A is not updated; otherwise no update occurs with the system state. The unitary operator for question A applied to the subspace for a type 1 individual is defined by
\begin{align*}
U_{A}\ket{0  0  1}\ket{0  0}  & =\ket{0  1  1}\ket{0  0} \\
U_{A}\ket{0  1  1}\ket{0  0}  & =\ket{0  1  1}\ket{0  1} \\
U_{A}\ket{1  0  1}\ket{0  0}  & =\ket{1  0  1}\ket{1  0} \\
U_{A}\ket{1  1  1}\ket{0  0}  & =\ket{1  0  1}\ket{1  1} 
\end{align*}
Likewise for question B, if the values on the belief state agree, then the value on belief A is changed to disagree with B and the belief of B is not updated; otherwise no update occurs with the system state.

Type 2 people update in the opposite way. They also copy the value on the belief states directly onto the response states, but again there is a back action that updates the system state as follows. For question A, if the values on the belief state disagree, then the value on belief B is updated to agree with A and the belief of A is not updated; otherwise no belief update occurs with the system state. The unitary for question A applied to the subspace for a type 2 individual is defined by
 \begin{align*}
U_{A}\ket{0  0  2}\ket{0  0}  & =\ket{0  0  2}\ket{0  0} \\
U_{A}\ket{0  1  2}\ket{0  0}  & =\ket{0  0  2}\ket{0  1} \\
U_{A}\ket{1  0  2}\ket{0  0}  & =\ket{1  1  2}\ket{1  0} \\
U_{A}\ket{1  1  2}\ket{0  0}  & =\ket{1  1  2}\ket{1  1} 
\end{align*}
Likewise for question B,  if the values on the belief state disagree, then the value on belief A is updated to agree with B and the belief of B is not updated; otherwise no update occurs with the system state. In sum, these unitary operators simply exchange one basis vector for another, and the system never becomes superposed with respect to the basis.

It is straightforward to explain how this model works to produce a belief update in the answers to A,B depending on question order. Suppose a type 1 person's initial state is $\ket{001}$. When this type of person is asked questions A and then B,  this person will answer no to A, and then update from the initial no state for B to answer yes to B. The probability for this to happen is based on the corresponding probability  $\mu_{001}$ of the initial state.  When this type of person is asked questions B and then A, this person will answer no to B, and then update from the initial no state for A to answer yes to A. This happens with the same probability  $\mu_{001}$ for the initial state.  Now suppose a type 2 person's initial state is $\ket{012}$. When this person is asked questions A and then B,  this person will answer no to A, and then update the state for B to answer no to B. This happens with probability $\mu_{012}$ for the initial state.  When this person is asked questions B and then A, this person will answer yes to B, and then update the A state to say yes to A. This happens with the same probability $\mu_{012}$ for the initial state.  In this way, the model produces question order effects that satisfy the $QQ$ equality in a parameter free manner.

Now consider answers to the second A question in the ABA paradigm. A type 1 person with initial state $\ket{001}$ will answer no to A, the state will update to $\ket{011}$ to say yes to question B, and then remain in the state $\ket{011}$ to say no to A the second time. A type 2 person with initial state  $\ket{102}$ will answer yes to A, the state will change to  $\ket{112}$ to say yes to question B, and then remain in the state $\ket{112}$ to say yes to A the second time.  In sum, this model always repeats the first answer to question A on the second repetition. It does not produce any changes in answers across the repetition. 

Formally, the model starts with an initial system state $\rho_{\bS}=\sum_{\alpha,\beta,\gamma} \mu_{\alpha,\beta,\gamma} \left|\alpha  \beta  \gamma\right\rangle \left\langle \alpha  \beta  \gamma \right|$, which is combined with the initial environment state $\rho_{\bE}=\left|0  0\right\rangle \left\langle 0  0\right|$  to produce the system plus environment state $\rho_{\bS+\bE}=\rho_{\bS}\otimes\rho_{\bE}$.  Then, the state  $\rho_{\bS+\bE}$  is transformed during measurement by the unitary operator and reduced by partial trace to an instrument.  For example, the instrument for the answer $x=0,1$ (1: yes, 0: no)  to question A is equal to 
$\mathcal{I}_{A}(x)\rho_{\bS}=\Tr_{\bE}\left[\left[I_{\bS}\otimes P^{M_A}(x)\right] U_{A}\left(\rho_{\bS}\otimes\rho_{\bE}\right) U_{A}^{\dagger}\right]$, 
with $P^{M_A}(x)=\ketbra{x}\otimes I_2$, where $\ketbra{x}$ acts on the 
2-dimensional space of the response to question A and 
$I_2$ is the identity operator on the 
2-dimensional space of  the response to question B.

Consider now asking question A and then B. The probability of yes to A and then no to B is determined from the product of instruments as follows. The probability of yes to A equals $p(A=1\|\rho_{\bS})=\Tr\left[\mathcal{I}_{A}(1)\rho_{\bS}\right]
=\Tr[P^{A}(1)\rho_{\bS}]$ with $P^{A}(1)=A=\ketbra{1}\otimes I_2\otimes I_3$, 
where $\ketbra{1}$ acts on the belief state space for question A, 
the identity operator $I_2$ acts on the  belief state space for question B, and
the identity operator $I_3$ acts on the personality state space.
Given that yes to A was observed, the belief state is updated to $\rho_{\bS\{A=1\}}=\frac{\mathcal{I}_{A}(1)\rho_{\bS}}{\Tr\left[\mathcal{I}_{A}(1)\rho_{\bS}\right]}$. Then the instrument for answering no to B after answering the question to A is defined as $\mathcal{I}_{B}(0)\rho_{\bS\{A=1\}}=\Tr_{\bE}\left[
\left[I_{\bS}\otimes P^{M_B}(0)\right]
 U_{B}\left(\rho_{\bS\{A=1\}}\otimes\rho_{\bE}\right) U_{B}^{\dagger}\right]$ 
 with $P^{M_B}(0)=I_2\otimes \ketbra{0}$
 and the probability of no to B given yes to A equals  
 $p(B=0|A=1\|\rho_{\bS})=\Tr\left[\mathcal{I}_{B}(0)\rho_{\bS\{A=1\}}\right]$. 
 Therefore, the sequential probability equals 
 $p(A=1,B=0\|\rho_{\bS}) =\Tr\left[\mathcal{I}_{B}(0)\mathcal{I}_{A}(1)\rho_{\bS}\right]$. 
 But, using the simple explanation described earlier, this turns out to equal to 
 $p(A=1,B=0)=\mu_{100}+\mu_{101}+\mu_{111}$ \cite[Eq.~(89)]{32}: 
 the probability of a type 0 person in state (100) that does not change the state, plus the probability of a type 1 person in state (101) that does not change this particular state, plus the probability of a type 1 person in state (111) that changes the state for B from yes to no. 

\subsection{System plus environment (BB) Model} 

The BB alternative account also assumes that the Hilbert space is composed of a tensor product of a system plus environment space. The system space is an $N$ - dimensional space used to represent a person's beliefs and opinions. A state vector within this space can be used to evaluate opinions about different questions depending on the basis used to evaluate the state.  

The environment is represented by a tensor product of a 2 dimensional space used to represent a person's response for question A the first time it is asked, another 2 dimensional space used to represent the person's response to question B, and another 2 dimensional space used to represent a person's response for question A the second time it is asked. The environment space is spanned by the 8 basis vectors  $\left\{ \left|0\right\rangle ,\left|1\right\rangle \right\} \otimes\left\{ \left|0\right\rangle ,\left|1\right\rangle \right\} \otimes\left\{ \left|0\right\rangle ,\left|1\right\rangle \right\}$. According to the BB model, the environment space represents the mental records of the measurements (i.e., episodic memories physically recorded in the neurons of the brain).

Define  $\left|+\right\rangle =(\frac{1}{\sqrt{2}})\left|0\right\rangle +(\frac{1}{\sqrt{2}})\left|1\right\rangle $, which represents an initial environmental state  with equal probabilities of responding yes or no to a question. Then the initial state is equal to the system plus environment state $\left|\zeta_{I}\right\rangle =\left|\psi_{I}\right\rangle \otimes(\left|+\right\rangle \otimes\left|+\right\rangle \otimes\left|+\right\rangle) $. A system state $\left|\psi\right\rangle $ can be decomposed using orthogonal projections $T_A,F_A=I-T_A$  for true and false opinions about question A: $\left|\psi\right\rangle =\left(T_{A}+F_{A}\right)\left|\psi\right\rangle =T_{A}\left|\psi\right\rangle +F_{A}\left|\psi\right\rangle $. Likewise, a system state can be decomposed using orthogonal projections $T_B,F_B=I-T_B$  for true and false opinions about question B: $\left|\psi\right\rangle  =T_{B}\left|\psi\right\rangle +F_{B}\left|\psi\right\rangle $. Four unitary operators are used to apply measurements to the questions. Suppose the questions are asked in the ABA order.

Intuitively, $U_A$ measures question A on the first occasion: if the person's belief state is experienced in the false subspace for question A, then the answer no is recorded in memory for this question; if the person's belief is experienced in the true subspace for question A, then the answer yes is recorded in memory for this question. 
\begin{align*}
U_{A}\left|\zeta_I\right\rangle   =\boldsymbol{F}_{A}\left|\psi_I\right\rangle \otimes\left|0\right\rangle \otimes\left|+\right\rangle \otimes\left|+\right\rangle +\boldsymbol{T}_{A}\left|\psi_I\right\rangle \otimes\left|1\right\rangle \otimes\left|+\right\rangle \otimes\left|+\right\rangle 
\end{align*}
Following the answer to question A, the system state reduces to $ \left|\psi_A\right\rangle $, which is the normalized state after observing the outcome $\left|i\right\rangle$ from question A. This state is then used to evaluate the next question about B.

Intuitively, $U_B$ measures question B: if the person's belief is experienced in the false subspace for question B, then the answer no is recorded in memory for this question; if the person's belief is experienced in the true subspace for question B, then the answer yes is recorded in memory for this question. 
\begin{align*}
U_{B}\left|\zeta_A\right\rangle   =\boldsymbol{F}_{B}\left|\psi_A\right\rangle \otimes\left|i\right\rangle \otimes\left|0\right\rangle \otimes\left|+\right\rangle +\boldsymbol{T}_{B}\left|\psi_A\right\rangle \otimes\left|i\right\rangle \otimes\left|1\right\rangle \otimes\left|+\right\rangle 
\end{align*}
Following the answer to question B, the state reduces to $ \left|\psi_{AB}\right\rangle$, which is the normalized state after observing the outcome  $\left|j\right\rangle$ from question B. This state is then used to evaluate the repeated question about A. 

The next two unitary operators correspond to the second instance of question A. It is assumed that there are two different types of participants, a "judgment based" type of responder and a "memory based" type of responder.\footnote{In \cite{34} we referred to the two types as conscientious and lazy. However these terms are overly narrow and restrictive. The so called lazy participants may be participants who thoughtfully answer questions A and B the first time they are presented, but then they copy their answer from the first time to the second time A is presented in order to save time and effort or to appear consistent in their responses.} Upon the second presentation of question A, the judgment based responder forms a new answer to the question "what is my opinion about question A." In contrast, the memory based responder answers the question "what did I answer the first time for question A."

The judgment based type of participant takes the time and effort to re-evaluate her opinion about question A when encountering it a second time and so makes a new judgment about question A, after evaluating B:
\begin{align*}
U_{J}\left|\zeta_{AB}\right\rangle   =\boldsymbol{F}_{A}\left|\psi_{AB}\right\rangle \otimes\left|i\right\rangle \otimes\left|j\right\rangle \otimes\left|0\right\rangle +\boldsymbol{T}_{A}\left|\psi_{AB}\right\rangle \otimes\left|i\right\rangle \otimes\left|j\right\rangle \otimes\left|1\right\rangle 
\end{align*}
The memory based participant simply copies the first answer from question A to the second: 
\begin{align*}
U_{M}\left|\zeta_{AB}\right\rangle  & =U_{M}\left|\psi_{AB}\right\rangle \otimes\left|i\right\rangle \otimes\left|j\right\rangle \otimes\left|+\right\rangle \\
 & =\left|\psi_{AB}\right\rangle \otimes\left|i\right\rangle \otimes\left|j\right\rangle \otimes\left|i\right\rangle .
\end{align*}

The probability that a participant chooses a response, such as yes, to a question is equal to the probability that the person observes that value in the episodic memory state corresponding to that question.  The measurement projections for computing probabilities of no versus yes to the first A question
are respectively 
\begin{align*}
\boldsymbol{A}_{F} & =I_N\otimes\left|0\right\rangle \left\langle 0\right|\otimes I_2\otimes I_2\\
\boldsymbol{A}_{T} & =I_N\otimes\left|1\right\rangle \left\langle 1\right|\otimes I_2\otimes I_2.
\end{align*}
Likewise the measurement projections for no versus yes to the second B 
question are respectively
\begin{align*}
\boldsymbol{B}_{F} & =I_N\otimes I_2\otimes\left|0\right\rangle \left\langle 0\right|\otimes I_2\\
\boldsymbol{B}_{T} & =I_N\otimes I_2\otimes\left|1\right\rangle \left\langle 1\right|\otimes I_2.
\end{align*}
Finally, the measurement projections for no versus yes to the third
question are respectively
\begin{align*}
\boldsymbol{A}'_{F} & =I_N\otimes I_2\otimes I_2\otimes\left|0\right\rangle \left\langle 0\right|\\
\boldsymbol{A}'_{T} & =I_N\otimes I_2\otimes I_2\otimes\left|1\right\rangle \left\langle 1\right|.
\end{align*}

If we measure question A and then B, we have to apply $U_{A}$ and then $U_{B}$ to the initial state and compute the probability of the relevant measurement outcomes. For example, consider the probability of answering yes to A first and then no to B second,
\begin{align*}
q(A=y,B=n) & =\left\Vert \boldsymbol{B}_{F}\boldsymbol{A}_{T}U_{B}U_{A}\left|\zeta\right\rangle \right\Vert ^{2}\\
 & =\left\Vert \left(\boldsymbol{F}_{B}\boldsymbol{T}_{A}\left|\psi\right\rangle \right)\otimes\left|1\right\rangle \otimes\left|0\right\rangle \otimes\left|+\right\rangle \right\Vert ^{2}\\
 & =\left\Vert \boldsymbol{F}_{B}\boldsymbol{T}_{A}\left|\psi\right\rangle \right\Vert ^{2}
\end{align*}
In the opposite order, we obtain
\begin{align*}
q(B & =n,A=y)=\left\Vert \boldsymbol{A}_{T}\boldsymbol{B}_{F}U_{A}U_{B}\left|\zeta\right\rangle \right\Vert ^{2}\\
 & =\left\Vert \left(\boldsymbol{T}_{A}\boldsymbol{F}_{B}\left|\psi\right\rangle \right)\otimes\left|1\right\rangle \otimes\left|0\right\rangle \otimes\left|+\right\rangle \right\Vert ^{2}\\
 & =\left\Vert \boldsymbol{T}_{A}\boldsymbol{F}_{B}\left|\psi\right\rangle \right\Vert ^{2}.
\end{align*}

Thus, for the first two A,B questions, the system plus environment model produces exactly the same sequential probabilities as the original question order model \cite{29} from which the $QQ$ equality was originally derived in a parameter free manner. It reproduces the repetition response for the AA type paradigm for both types of responders -- for the judgment-based responder, this follows from the properties of projection operators; for the memory-based responder, this follows from simple recall of memory. The BB model also reproduces the $QQ$ equality for the AB,BA type paradigm. This follows from the fact that there are no repeated questions in this paradigm, and so both type of responders must rely on judgments using the projection operators. For the third question, the repetition of question A, the judgment-based responder (using $U_J$ to record a response to the repetition to A) can produce a different answer compared to what they originally responded to A (because they re-evaluate the question). However, a memory- based responder  (using $U_M$ to record a response to the repetition to A) will simply repeat their previous answer. 

\subsection{New (OB) model} 
During discussions of the previous two models, the authors developed a third model (OB model) that shares properties of the first two. It shares the measurement properties of the OK model and the repeat answer property of the BB model. 

The mental state, or the system state to be measured,  
consists of a belief state in a 2 dimensional Hilbert space
$\cH_{\bf b}$, and a memory state in a $3 \cdot 3$ dimensional Hilbert space $\cH_{\bf m}$,
so that the state space $\cH_{\bS}$ of the system $\bS$ to be measured is their tensor product $\cH_{\bS}=\cH_{\bf b}\otimes\cH_{\bf m}$.
The state space of the environment $\bE$ recording the responses to a single instrument
is a $3\cdot 2$ dimensional  Hilbert space $\cH_{\bE}$. Altogether the model uses a $2 \cdot 9 \cdot 6$ dimensional space $\cH_{\bS}\otimes \cH_{\bE}=\cH_{\bf b} \otimes \cH_{\bf m} \otimes \cH_{\bE}$ describing the measuring process of a single instrument.\footnote{For a sequence $\cI_1,
\cdots,\cI_n$ of instruments measuring the same system $\bS$, the measuring process of the
whole sequential measurements is described on the tensor product space $\cH_{\bS}\otimes \cH_{\bE_1}\otimes \cdots \otimes \cH_{\bE_n}$,
where $\cH_{\bE_j}$ represents the environment $\bE_j$ of the instruments $\cI_j$.
However, once the instruments $\cI_j$ is determined by the system-environment description on 
$\cH_{\bS}\otimes\cH_{\bE_j}$, the sequential probability of their  outcomes 
with the initial system state $\rh$ is given
by $p(\bx_1=x_1,\ldots,\bx_n=x_n\|\rh)=\Tr[\cI_n(x_n)\cdots\cI_n(x_1)\rh]$.}

The belief states in $\cH_{\bf b}$ use two different bases:
a basis $\left\{ \left|O_{A}\right\rangle ,\left|1_{A}\right\rangle \right\} $
for question A, and a basis $\left\{ \left|O_{B}\right\rangle ,\left|1_{B}\right\rangle \right\} $
for question B. The memory state in $\cH_{\bf m}$ has 9 basis vectors $\left\{ \left|xy\right\rangle ,x=0,1,2;y=0,1,2\right\} $
with $x$ representing the mental state for question A and $y$ representing
the mental state for question B; the index 0 represents memory for
a previous no/false answer to the question, 1 represents a memory for
a previous yes/true answer to the question, and the index 2 represents
no previous answer to the question. The response state in $\cH_{\bE}$ has six basis vectors $\left\{ \left|uv\right\rangle ,u=0,1,2;v=0,1\right\} $, where $v=0$ indicates a response of no/false, and $v=1$ indicates a response of yes/true, and the index $u=0,1,2$ is used to represent the memory state for the question being asked.

Then the unitary operator $U_{A}$ is defined as 
\begin{align*}
\left|a_{A}\right\rangle \left|xy\right\rangle \left|00\right\rangle  & \rightarrow\left|a_{A}\right\rangle \left|f(x)y\right\rangle \left|xf(x)\right\rangle ,
\end{align*}
with $a$ taking on values 0 or 1 and $f(0)=0,$ $f(1)=1,$ and $f(2)=a.$ Similarly, the unitary operator
$U_{B}$ is defined as 

\begin{align*}
\left|b_{B}\right\rangle \left|xy\right\rangle \left|00\right\rangle  & \rightarrow\left|b_{B}\right\rangle \left|xg(y)\right\rangle \left|yg(y)\right\rangle ,
\end{align*}
with $b$ taking on values of 0 or 1 and $g(0)=0,g(1)=1,g(2)=b.$ 

The initial state, before any questions have been asked, is defined as 
\[
\left|\psi(0)\right\rangle =\left|\\ \gamma\right\rangle \left|22\right\rangle \left|00\right\rangle, 
\]
where  $\left|\gamma\right\rangle $ is a vector in $\cH_{\bf b}$ representing the person's beliefs. The initial density matrix is then equal to $\rho_{0}=\left|\psi_{0}\right\rangle \left\langle \psi_{0}\right|$.

Note that in both cases, if there is
a previous memory state, then it is repeated in the response.
If there is no previous memory for the question, the response is set
equal to the current belief state. In the latter case, the change in basis used to represent the beliefs to each question produces question order effects. 

The probability that the participant chooses a response is determined by the second component of 
the response state in $\cH_{\bE}$. Define the projection for detecting a yes/true probe response as $M_{y}=I_{3}\otimes\left|1\right\rangle \left\langle 1\right|$, where $I_3$ is a $3 \times 3$ identity matrix. Likewise, the projection for a no/false response is defined by 
$M_{n}=I_{3}\otimes\left|0\right\rangle \left\langle 0\right|$.

The instrument for a true/yes/1 response to question A is given by the partial trace of the projection on yes over the environment states after the unitary evolution $U_A$
\[
\mathcal{I}_{A}(1)\rho_{\bS}=\Tr_{\bE}\left[\left(I_{2}\otimes I_{9}\otimes M_{y}\right)U_{A}  (\rho_{\bS}\otimes\rho_{\bE})  U_{A}^{\dagger}\right],
\]
where $I_n$ is a $n \times n$ identity matrix, and $\rho_{\bE}$ is the initial environment state 
set as $\rho_{\bE}=\ket{00}\bra{00}$,
and similarly question B
\[
\mathcal{I}_{B}(1)\rho_{\bS}=\Tr_{\bE}\left[\left(I_{2}\otimes I_{9}\otimes M_{y}\right)U_{B}  (\rho_{\bS}\otimes \rho_{\bE})  U_{B}^{\dagger}\right],
\]

Then the probability of a sequence, for example answer $x$ to A then $y$ to B then $z$ to A can be computed from 

\[
p(A=x,B=y,A=z\|\rho_{\bS})=\Tr\left[\mathcal{I}_{A}(z)\mathcal{I}_{B}(y)\mathcal{I}_{A}(x)\rho_{\bS}\right].
\]

The new OB model produces question order effects in the same way as the original order model \cite{29} and it repeats the answer to question A based on memory for the first answer to A like the BB model, but it uses instruments to compute the probabilities of {\em any} sequences of events like the OK model.

\subsection{Comparison}

There are several important differences between the OK, BB, and OB models that are useful to discuss. 

First,  there is a difference concerning the compatibility nature of the two questions A,B.  The OK account uses a common tensor product basis to represent both questions. This assumes that a person can simultaneously hold a firm belief on both questions -- accordingly, the two questions are assumed to be compatible. In contrast, the BB and OB accounts use a different basis to represent the answers for each question -- consequently, the two questions are assumed to be incompatible. However, for the BB and OB models, the memories for the answers to the question are represented in a tensor product state that assumes the person can simultaneously remember the previous answers to the questions.  

A benefit of using a tensor product belief state is that the OK model can provide an estimate of the initial joint probability distribution of beliefs across the questions before being disturbed by measurement. The BB and OB models assume that no  joint prior distribution even exists.   Behaviorally, we can imagine that certain types of stimuli can be processed either sequentially or simultaneously, depending on task demands. However,  sequential processing seems to be an unavoidable requirement for many cases of decision making \cite{35};
even if a model assumes a joint probability for prior beliefs for both questions, the model describes 
how the process of answering to one questions affect the belief on the other question.    

A second important difference concerns the measurement process that produces the question order effect in the AB,BA paradigm.  According to the OK model, the question order effect occurs because a unitary operator directly changes the beliefs about the questions after measurement. For \textit{any} pair of questions, a type I person changes an initial belief that both answers agree to a belief that the answers disagree; the type II person changes an initial belief that the answers disagree to a belief that the answers agree.  In contrast, according to the BB and OB models, the question order effect occurs because the projections used to evaluate the two questions do not commute. This principle led to the \textit{a priori} prediction of the $QQ$ equality. This same principle has been used to account for many other puzzling findings in the judgment and decision making literature \cite{36}.

Third, there is a difference concerning the assumptions about the mapping of experimental questions to measurement operations. The OK and OB models apply one type of measurement operation (or instrument, more precisely) for each experimentally presented question. In particular, the same operation for question A is used regardless of whether A is presented the first time or second time. The BB model introduces a new measurement operation for the memory-based participants on the second repeated presentation.  Thus, even though the experimenter repeats the question  "do you agree with A," the memory based participant answers a different question "what did I answer about A the first time."  Thus, the BB model does not require the participant to evaluate the experimenter's question. In other words, the BB model assumes that the experimenter is not in complete control of what the participant is thinking. 

For the above reason, the BB model needs to add an additional memory recall operator when a question is repeated. On the one hand, this addition of a memory recall operator is more complex than the simpler pair of instruments used by OK. On the other hand, as pointed out in \cite{28}, recalling previous responses on repeated questions is an old and well known problem in experimental psychology. Experimental psychologists generally go to great extents using filler items in between repetitions in order to try to avoid the possibility of the participant simply recalling previous answers (see e.g., \cite{37}).  

Another phenomena that can occur with the BB and OB models is that when answering the sequence of questions ABA, the  memory recorded from the first presentation of question A may no longer be completely consistent with the belief state that results after answering question B. This follows from the disturbance produced by the non-commuting projections. So, for example, a person may remember that they initially said that they agree with a statement A, but now experience some doubt about their opinion about statement A. This inconsistency of memory and beliefs doesn't arise with the OK model because the instruments used by the OK model retain memory of the probe responses in the belief state. For example, following the measurement of A, a type I person initially in state  $\ket{111}$ updates to $\ket{101}$ that has an environment state $\ket{11}=\ket{1}\ket{1}$, in which the first qubit $\ket{1}$ shows the measurement outcome, while the second qubit $\ket{1}$
is the dummy to ensure the unitarity of $U_A$. In this model, the memory of the previous outcome is kept in the first qubit $\ket{1}$ of the updated belief state $\ket{101}$; in this sense, the memory and the belief are   
unified in this model.

Finally, the OK and OB accounts require all participants to repeat their answers from the first to the second presentation of question A (with question B answered in between), whereas the BB account allows some participants to repeat while others may change their mind. Empirically the latter has been observed (see, \cite{30}), but these results are controversial because participants were instructed to reconsider their answers to question A after evaluating question B. More research is clearly required to determine the empirical findings for the ABA paradigm. The OK account could be generalized to include additional types and thereby allow changes in answers to the repeated question. However, this modification would need to be made in a way that is consistent with the empirical results related to the $QQ$ equality in a parameter free manner.

\section{Conclusion}

Both Bayesian and quantum models have been proposed to account for the human judgments and reasoning under uncertainty. An important challenge for these theories is to address the effects of episodic memories on judgments and decisions. For example, how does making a prediction about an event affect later decisions about actions. One of the main advantages of quantum models over Bayesian models is their capability to account for the episodic memory effects of recently made measurements on later judgments and decisions, such as question order effects. 

This article  reviews new progress on quantum models that address this issue by generalizing the type of measurement operations used in quantum cognition. The early quantum measurement models were based on projective ``collapse'' type measurement models that quickly lose memory, as each new question is asked. In this article, we describe more modern measurement models based on a system plus environment representation of the state, and the use of generalized instruments to describe the measurement operations. We applied these more modern measurement models to three empirical paradigms concerning question order effects in human judgment research.

The first quantum question order model \cite{29} used non-commuting projections to provide the original basis for predicting both the response replication effect for the AA repeated question paradigms, and the $QQ$ equality for the AB,BA paradigm. However, as pointed out by \cite{30}, it fails to account for those participants who repeat their answers to A on the ABA paradigm.  This theoretical problem was first resolved by \cite{32} who proposed using instruments instead of non-commuting projections to account for question order effects. Later,  the system plus environment model 
 was proposed by \cite{33}
that uses non-commuting projections as in the original question order model, but introduced a type of individual that copies answers from the first time question A is asked to the second time to account for those participants who repeat their answers to A on the ABA paradigm. 

Note that the difference between the OK and BB models does not depend on whether or not instruments are ultimately used. An instrument version of the BB model could be constructed just as the OB model did by expanding the Hilbert space to include environment states. Then control-U gates could be used to copy the values in the memory states onto the environment states. Instruments could then be constructed by marginalizing over the environment state as in the OK model. This would  produce the same final choice probabilities as the original BB model.  Therefore, the difference between models arises from how episodic memory for the previous responses are used to make future responses.

Despite the differences between the OK, BB, and OB models, they all assume that measurement can change a person's beliefs, the change depends on the order of measurement, and that  quantum models of question order effects need to include the effects of episodic memory from past judgments.

We have only considered one kind of memory in this article, memory for experiences, called episodic memory. Semantic memory is another kind of memory, corresponding to meaning extracted across many experiences. In quantum cognition, there are two approaches for representing the meaning of a concept, like dog or cat. According to the State Concept Property system  \cite{38}, concepts are represented as vectors in a Hilbert space modified by context. Alternatively, the meaning of a concept  can be represented as multi-dimensional projections that contain basis states representing the features describing the concept \cite{39,40}. However, this is a complex topic that needs to be addressed in future research.  

\ack{We thank Andrei Khrennikov and Peter Bruza for past contributions and discussions on these issues.
}
\funding{M.O. was partially supported by JSPS KAKENHI Grant Numbers JP25K07108,
JP24H01566, JP22K03424,  JST CREST Grant Number JPMJCR23P4, 
and the Quantinuum--Chubu University Collaboration in 2023--2024.
E.M.P. was supported by European Office of Aerospace Research and Development (EOARD) grant FA8655-23-1-7220.}.

%%%%%%%%%% Insert bibliography here %%%%%%%%%%%%%%


\begin{thebibliography}{29}

\bibitem{1} Griffiths TL, Chater N, Tenenbaum JB.  2024  \textit{Bayesian models of cognition: reverse engineering the mind}. 
Cambridge, MA: MIT Press.

\bibitem{2} Busemeyer JR, Bruza PD. 2012 \textit{Quantum models of cognition and decision.} Cambridge, UK: Cambridge University Press.

\bibitem{3} Khrennikov A.  2010 \textit{Ubiquitous quantum structure}. Berlin: Springer. 

\bibitem{4}  Pothos EM, Busemeyer JR. 2022 Quantum cognition.  \textit{Annual review of psychology} {\bf 73}, 749--778.

\bibitem{5} Malmberg K. 2024 \textit{Human Memory: the general theory and its various models}. Cambridge, UK: Cambridge University Press.

\bibitem{6} Murdock BB. 1974 \textit{Human memory: Theory and data}. Lawrence Erlbaum.

\bibitem{7} Denolf J,  Lambert-Mogiliansky A. 2016 Bohr complementarity in memory retrieval. \textit{Journal of Mathematical Psychology}, 73:28--36.

\bibitem{8} Trueblood JS, Hemmer P. 2017 The generalized quantum episodic memory model. \textit{Cognitive science} {\bf 41}, 2089--2125.

\bibitem{9}von Neumann J.  1955  \textit{Mathematical Foundations of Quantum  Mechanics}. Princeton University Press [Originally published: J. von Neumann. 1932 \textit{Mathematische Grundlagen der Quantenmechanik}. Berlin: Springer].

\bibitem{10}  Yearsley JM,  Pothos EM.  2016 Zeno's paradox in decision-making.  \textit{Proceedings of the Royal Society B: Biological Sciences} {\bf 283}, 20160291.

\bibitem{11} Dirac PAM. 1958 \textit{The Principles of Quantum Mechanics, 4th edition.} Oxford UK: Oxford University Press.

\bibitem{12} Schr\"{o}dinger E. 1935  Die gegenw\"{a}rtige Situation in der Quantenmechanik. \textit{Naturwissenshaften} {\bf 23}, 807--812, 823--828, 844--849. [English translation: JD Trimmer, 1980 {\em Proc.\ Am.\ Philos.\ Soc.} {\bf 124}, 323--338.]

\bibitem{13} L\"{u}ders G. 1950 \"{U}ber die Zustands\"{a}nderung durch den Me{\ss}proze{\ss}. \textit{Annalen der Physik} {\bf 443}, 322--328.

\bibitem{14} Kvam PD, Pleskac TJ, Yu S, Busemeyer JR. 2015 Interference effects of choice on confidence: Quantum characteristics of evidence accumulation. \textit{Proceedings of the National Academy of Sciences} {\bf 112}, 10645--10650.

\bibitem{15} Ozawa M, Khrennikov A. 2023  Nondistributivity of human logic and violation of response replicability effect in cognitive psychology.
\textit{Journal of Mathematical Psychology} {\bf 112}, 102739.

\bibitem{16} Davies EB, Lewis JT. 1970 An operational approach to quantum probability. \textit{Communications in Mathematical Physics} {\bf 17}, 239--260. 

\bibitem{17} Ozawa M. 2000  Measurements of nondegenerate discrete observables. {\em Physical Review A} {\bf 62}, 062101.

\bibitem{18} Ozawa M. 2004 Uncertainty relations for noise and disturbance in generalized  quantum measurements. {\em Annals of  Physics} {\bf 311}, 350--416.

\bibitem{19} Ozawa M. 2014 Mathematical foundations of quantum information: {Measurement} and foundations. {\em Sugaku Expositions} {\bf 27}, 
195--221.  (https://arxiv.org/abs/1201.5334)

\bibitem{20}Ozawa M. 1984 Quantum measuring processes of continuous observables.  \textit{Journal of Mathematical Physics} {\bf 25}, 79--87.

\bibitem{21} Ozawa M. 1997 An operational approach to quantum state reduction. \textit{Annals of Physics} {\bf 259}, 121--137. 

\bibitem{22} Ozawa M. 2023 Quantum measurement theory for systems with finite dimensional state spaces. In \textit{The Quantum-Like Revolution: A Festschrift for Andrei Khrennikov} (eds A Plotnitsky, E Haven),  pp.~191--214. Cham, Switzerland: Springer International Publishing.

\bibitem{23} Ozawa M. 1997 Quantum state reduction and the quantum {Bayes} principle. In {\em Quantum  Communication, Computing, and Measurement} (eds O Hirota, AS Holevo, CM Caves), pp.~233--241. New York, NY: Plenum. 

\bibitem{24} Ozawa M. 1998 Quantum state reduction: {An} operational approach. {\em Fortschritte der Physik} {\bf 46}, 615--625.

\bibitem{25} Ozawa M. 2001 Operations, disturbance, and simultaneous measurability. {\em Physical Review A} {\bf 63}, 032109.

\bibitem{26} Nielsen MA, Chuang IL. 2000 \textit{Quantum computation and quantum information.} Cambridge UK: Cambridge University Press.

\bibitem{27} Kraus K. 1971 General state changes in quantum theory. \textit{Annalen der Physik} {\bf 64}, 311--335.

\bibitem{28} Khrennikov A, Basieva I, Dzhafarov EN, Busemeyer JR. 2014 Quantum models for psychological measurements: an unsolved problem. \textit{PloS One} {\bf 9}, e110909.

\bibitem{29} Wang Z, Solloway T, Shiffrin RM, Busemeyer JR. 2014 Context effects produced by question orders reveal quantum nature of human judgments. \textit{Proceedings of the National Academy of Sciences} {\bf 111}, 9431--9436.

\bibitem{30} Busemeyer JR, Wang Z.  2017 Is there a problem with quantum models of psychological measurements?. \textit{PloS One} {\bf  12}, e0187733.

\bibitem{31} Wang Z, Busemeyer JR. 2013  A quantum question order model supported by empirical tests of an a priori and precise prediction. {\em Topics in Cognitive Science} {\bf  5}, 689--710.

\bibitem{32} Ozawa M, Khrennikov A. 2021 Modeling combination of question order effect, response replicability effect, and QQ-equality with quantum instruments. \textit{Journal of Mathematical Psychology} {\bf 100}, 102491. 

\bibitem{33} Busemeyer JR. 2023 Measurement Models in Quantum Cognition. In \textit{The Quantum-Like Revolution: A Festschrift for Andrei Khrennikov}  (eds A Plotnitsky, E Haven), pp.~269--279. Cham, Switzerland: Springer International Publishing.

\bibitem{34} Busemeyer JR, Bruza PD. 2024  \textit{Quantum models of cognition and decision. 2nd Edition.} Cambridge  UK: Cambridge University Press.

\bibitem{35}  Epping G, Fisher EL, Zeleznikov-Johnston AM, Pothos EM, Tsuchiya N. 2023 A quantum geometric framework for modeling color similarity judgments. \textit{Cognitive Science} {\bf 47}, e13231.

\bibitem{36} Pothos EM, Busemeyer JR. 2022 Quantum cognition. \textit{Annual Review of Psychology} {\bf 73}, 749--778.

\bibitem{37} Regenwetter M, Dana J, Davis-Stober CP. 2011 Transitivity of preferences. \textit{Psychological Review} {\bf 118}, 42--56.

\bibitem{38} Aerts D. 2000. Quantum structure in cognition. \textit{Journal of Mathematical Psychology} {\bf 53}, 314--348.

\bibitem{39} Busemeyer JR, Pothos EM, Franco R, Trueblood JS. 2011 A quantum theoretical explanation for probability judgment errors. \textit{Psychological Review} {\bf 118}, 193--218.

\bibitem{40} Pothos EM, Busemeyer JR,  Trueblood JS. 2013 A quantum geometric model of similarity. \textit{Psychological Review} {\bf 120}, 679--696.

\end{thebibliography}
\end{document}